 \documentclass[final,5p,times,twocolumn]{elsarticle}

 \usepackage{hyperref}

\usepackage{amssymb}
\usepackage{lipsum}
\usepackage{amsmath}

\makeatletter
\def\ps@pprintTitle{%
 \let\@oddhead\@empty
 \let\@evenhead\@empty
 \let\@oddfoot\@empty
 \let\@evenfoot\@empty}
\makeatother

\journal{Physics Letters B}

\begin{document}

\begin{frontmatter}



\title{Boosted dark matter from semi-annihilations in the galactic center}


\author[first,second]{Boris Betancourt Kamenetskaia}
\ead{boris.betancourt@tum.de}
\author[first,third]{Motoko Fujiwara}
\ead{motoko@sci.u-toyama.ac.jp}
\author[first]{Alejandro Ibarra}
\ead{ibarra@tum.de}
\author[fourth,fifth]{Takashi Toma}
\ead{toma@staff.kanazawa-u.ac.jp}
\affiliation[first]{organization={Technical University of Munich, TUM School of Natural Sciences, Physics Department, Professorship T30d Elementary Particle Physics},
            addressline={James-Franck-Str. 1}, 
            city={Garching},
            postcode={85748}, 
            country={Germany}}
\affiliation[second]{organization={Max-Planck-Institut für Physik (Werner-Heisenberg-Institut)},
            addressline={Boltzmannstra\ss e 8}, 
            city={Garching},
            postcode={85748}, 
            country={Germany}}
\affiliation[third]{organization={University of Toyama, Department of Physics},
            addressline={3190 Gofuku}, 
            city={Toyama},
            postcode={930-8555}, 
            country={Japan}}
\affiliation[fourth]{organization={Institute for Theoretical Physics, Kanazawa University},
            city={Kanazawa},
            postcode={920-1192}, 
            country={Japan}}
\affiliation[fifth]{organization={Institute of Liberal Arts and Science, Kanazawa University},
            city={Kanazawa},
            postcode={920-1192}, 
            country={Japan}}

\begin{abstract}
In some scenarios, the dark matter relic abundance is set by the semi-annihilation of two dark matter particles into one dark matter particle and one Standard Model particle. These semi-annihilations might still be occurring today in the Galactic Center at a significant rate, generating a flux of boosted dark matter particles. We investigate the possible signals of this flux component in direct detection and neutrino experiments for sub-GeV dark matter masses. We show that for typical values of the semi-annihilation cross-section, the sensitivity of current experiments to the spin-independent dark matter-proton scattering cross-section can be several orders of magnitude larger than current constraints from cosmic-ray boosted dark matter. We also argue that the upcoming DARWIN and DUNE experiments may probe scattering cross-sections as low as $10^{-37}\,{\rm cm}^2$ for masses between 30 MeV and 1 GeV.
\end{abstract}



\end{frontmatter}




\section{Introduction}
\label{introduction}

The freeze-out mechanism stands out as one of the most plausible mechanisms to explain the origin of a population of dark matter (DM) particles in our Universe. This framework assumes that DM particles were in thermal equilibrium with the plasma of Standard Model particles at very early times. The same interactions that ensure the thermalization of the DM particles with the Standard Model particles would lead to an exponential depletion of the yield of DM particles when the temperature drops below the DM mass, $m_{\chi}$, which stops when the expansion rate of the Universe becomes larger than the annihilation rate. After this epoch, the yield of DM particles ``freezes-out" and remains approximately constant until today, constituting a source of space-time curvature in galaxies, clusters of galaxies, and the Universe at large scale.

This appealing mechanism requires i) the thermalization of the DM particles with the Standard Model particles at temperatures much larger than the DM mass, and ii) a depletion of their yield from their initial thermal value to the observed value. It is important to notice that the latter ingredient does not necessarily require the annihilation of two DM particles into two Standard Model particles, but any process converting $n$ DM particles into $m$ DM particles, $n\rightarrow m$ with $n>m$, possibly accompanied by other Standard Model particles. While the simplest alternative is the $2\rightarrow 0$ process, with no DM particles in the final state, this possibility is not unique. Notable examples are the processes $3\rightarrow 2$ or $4\rightarrow 2$ ~\cite{Carlson:1992fn,PhysRevLett.113.171301} (see also ~\cite{Bernal:2017mqb,Ho:2017fte,Hochberg:2018rjs,Herms:2018ajr, Arcadi:2019oxh,Smirnov:2020zwf}). Parametrizing the thermally averaged cross-sections by $\langle \sigma v^{n-1}\rangle=\alpha_{\rm eff}^n/m_{\chi}^{3n-4}$, $n=3,4$, it can be shown that the freeze-out of these processes can reproduce the observed DM abundance if the DM field has strong self-interactions, $\alpha_{\rm eff}=1$, and the DM mass is $m_{\chi}\sim 40$ MeV or 100 keV respectively. A second example is the DM semi-annihilation $2\rightarrow 1$~\cite{Hambye:2008bq,Hambye:2009fg,Arina:2009uq,DEramo:2010keq}, which is necessarily accompanied by some Standard Model particles to ensure the conservation of energy-momentum. 

The processes $n\rightarrow m$, $n>m$,  convert part of the mass of the initial DM particles into kinetic energy of the final DM particles, thereby constituting a source of ``boosted" DM (other mechanisms that generate a flux of DM particles with velocities larger than those expected in the Standard Halo Model are the cosmic-ray boosted DM~\cite{Yin:2018yjn,Bringmann:2018cvk,Ema:2018bih}, annihilation/decay of heavy DM particles \citep{Agashe:2014yua,Kopp:2015bfa}  blazar-boosted, DM~\cite{Wang:2021jic}, or solar reflection of DM~\cite{Emken:2021lgc}). In this paper, we focus on the possible signatures of semi-annihilation boosted DM,  assuming that DM particles also interact with the proton. Signals of this process from capture in the Sun, and the subsequent semi-annihilation $\chi\chi\rightarrow\chi\nu$ have been studied in~\cite{Toma:2021vlw, Aoki:2023tlb}. The non-observation of the boosted DM from the Sun leads to upper limits on the DM-nucleon cross-section for DM masses above $4~\mathrm{GeV}$. For smaller DM masses, the DM evaporation due to thermal effects inside the Sun prevents the existence of a significant overdensity of DM. Therefore, the semi-annihilation signals become very suppressed. For light DM, hence, other targets must be considered. In this letter, we study the signals of semi-annihilation boosted DM in the Galactic Center, either in direct detection experiments (e.g. XENONnT \citep{XENON:2023cxc}, CRESST-II \citep{CRESST:2015txj}) or in neutrino detectors (e.g. MiniBooNE \citep{MiniBooNE:2008paa}), and we discuss the complementarity of the constraints and sensitivities to other search strategies.

This work is organized as follows:
In Sec.~\ref{sec:direct_detection} we review the formalism to calculate the event rate at direct detection experiments, extending to the possibility when the scattering is not purely coherent.  
In Sec.~\ref{sec:direct_detection_constraints}, we derive the constraints on this scenario from direct detection experiments, as well as the prospects for future experiments.
Finally, in Sec.~\ref{sec:conclusions} we present our conclusions. We also include Appendix~\ref{app:derivation}, which provides details on the calculation of the limits on the cross-section from the experimental data.

\section{Differential scattering rate}
\label{sec:direct_detection}

We assume that the Milky Way is embedded in a halo of DM particles with density distribution described by a Navarro-Frenk-White profile~\cite{Navarro:1995iw,Navarro:1996gj}:
\begin{align}
  \rho_{\rm  NFW}  (r)  &= \frac{\rho_s}{\left(\frac{r}{r_s} \right)  \left( 1  +  \frac{r}{r_s} \right)^2},
\end{align}  
with $\rho_s  =  0.184~\mathrm{GeV}/\mathrm{cm}^3$ and $r_s  =  24.42~\mathrm{kpc}$~\cite{Bergstrom:1997fj,Turner:1985si,Bertone:2004pz,Cirelli:2010xx}. We consider for concreteness the semi-annihilation of a spin 1/2 DM particle $\chi\chi\rightarrow \chi \nu$, which produces a flux of DM particles at the location of the Solar System given by
\begin{equation}
    \frac{d\Phi}{dT_\chi}=\Phi_{\rm BDM}\delta\Big(T_\chi-\frac{m_\chi}{4}\Big),
\label{eq:dPhi/dT_BDM}
\end{equation}
with
\begin{equation}\label{eq:Phi_BDM}
  \Phi_{\rm  BDM}  
  =  
  \frac{1}{2}  \frac{1}{4  \pi}  
    \int  d  \Omega  \int_{\rm  l.o.s}  d  s
  \left( \frac{\rho_\chi  (r ( s,  \theta ) )}{m_\chi} \right)^2\langle{\sigma_{2  \to  1}  v}\rangle,
\end{equation}
where $r  ( s,  \theta )  =  \sqrt{r_\odot^2  +  s^2  -  2  r_\odot  s  \cos  \theta}$ and  $r_\odot  =  8.33~\mathrm{kpc}$ is
the distance of the Sun to the galactic center.  Since current direct search experiments have no directional information, we will calculate the flux integrating the solid angle in Eq.~(\ref{eq:Phi_BDM}) over the whole sky.  Numerically, we obtain:
\begin{align}
\Phi_{\rm BDM}\simeq 3.2\times10^{-3}~\mathrm{cm}^{-2}~\mathrm{s}^{-1}~ 
\left(\frac{m_\chi}{100\,{\rm MeV}}\right)^{-2}
\left(\frac{\langle\sigma_{2\rightarrow 1} v\rangle}{10^{-26}\,{\rm cm}^{3}{\rm s}^{-1}}\right)\;.
\label{eq:num_flux}
\end{align}
The semi-annihilation produces an identical flux of neutrinos, with energy $T_\nu=3 m_\chi/4$, which would be equivalent to the signal produced in the annihilation $\psi\psi\rightarrow \nu\bar\nu$ with $m_\psi=3m_\chi/4$. The neutrino flux is undetectable with current instruments, which only exclude an annihilation cross-section for annihilation into $\nu\bar\nu$ larger than $\sim 10^{-24}\,{\rm cm}^3\,{\rm s}^{-1}$~\cite{Super-Kamiokande:2020sgt}. On the other hand, the flux of DM particles may induce a detectable number of nuclear recoil events in a direct detection experiment. The differential rate per target mass of scattering of DM with the nucleus ${\cal T}$, producing a recoil of the latter with kinetic energy $T_{\cal T}$, is given by:
\begin{equation}
    \frac{dR_{\cal T}}{dT_{\cal T}}=\frac{1}{m_{\cal T}}\int\displaylimits_{T_{\chi,{\cal T}}^{\rm min}(T_{\cal T})}^{\infty}{dT_\chi\,\frac{d\sigma_{\chi {\cal T}}}{dT_{\cal T}}(T_\chi,T_{\cal T})\frac{d\Phi}{dT_\chi}}.
    \label{eq:diff_rate}
\end{equation}
Here,  $m_{\cal T}$ is the mass of the target nucleus,  $T_\chi$ is the kinetic energy of the incoming DM particle, and $T_{\chi,{\cal T}}^{\rm min}$ is the minimal DM kinetic energy capable of producing in the final state a nucleus ${\cal T}$ recoiling with kinetic energy $T_{\cal T}$, which is given by: 
\begin{equation}\label{eq:T_min}
    T_{\chi,{\cal T}}^{\min}(T_{\cal T})=\left(\frac{T_{\cal T}}{2}-m_\chi\right)\left[1\pm\sqrt{1+\frac{2T_{\cal T}}{m_{\cal T}}\frac{(m_\chi+m_{\cal T})^2}{(T_{\cal T}-2m_\chi)^2}}\right],
\end{equation}
with the $+$ ($-$) for $T_{\cal T}>2m_\chi$ ($T_{\cal T}<2m_\chi$).  

Additionally, $d\sigma_{\chi{\cal T}}/dT_{\cal T}$ is the differential scattering cross-section for the scattering of a DM particle with the nucleus ${\cal T}$. It is important to note that DM particles from semi-annihilations have sizable speeds (see Eq.~(\ref{eq:dPhi/dT_BDM})). Therefore, for sufficiently large masses, the inverse of the momentum transfer could be comparable to or smaller than the size of the target nucleus, which translates into a loss of coherence of the scattering with the whole nucleus. We model this loss of coherence following \cite{Bednyakov:2018mjd,Bednyakov:2021bty} so that the total differential cross-section is expressed as a sum of a coherent and an incoherent part:
\begin{align}
 \frac{d\sigma_{\chi{\cal T}}}{dT_{\cal T}}= \left(\frac{d\sigma_{\chi{\cal T}}}{dT_{\cal T}}\right)_{\rm coh}+
 \left(\frac{d\sigma_{\chi{\cal T}}}{dT_{\cal T}}\right)_{\rm inc},
 \label{eq:diff_cross_section}
\end{align}
which for spin-independent interaction reads explicitly:
\begin{align}\label{eq:diff_scat_sigma}
    \left(\frac{d\sigma_{\chi{\cal T}}}{dT_{\cal T}}\right)_{\rm coh}&=
    \frac{ \sigma^{\rm coh}_{{\rm SI},{\cal T}}}{ T_{\cal T}^{\rm max}}|F_{{\rm SI},{\cal T}}(q)|^2,\\
    \left(\frac{d\sigma_{\chi{\cal T}}}{dT_{\cal T}}\right)_{\rm inc}&=
    \frac{ \sigma^{\rm inc}_{{\rm SI},{\cal T}}}{ T_{\cal T}^{\rm max}}\left(1-|F_{{\rm SI},{\cal T}}(q)|^2\right).
\end{align}
 This formula provides a smooth transition between the coherent and incoherent regimes. Here
\begin{equation}
T_{\cal T}^{\rm max}=\frac{T_\chi^2+2m_\chi T_\chi}{T_\chi+(m_\chi+m_{\cal T})^2/(2m_{\cal T})},
\label{eq:T_max}
\end{equation}
is the maximum kinetic energy of the target particle after scattering, and $q=\sqrt{2m_{\cal T}T_{\cal T}}$ is the momentum transfer. Moreover, $F_{\rm SI,{\cal T}}(q)$ is the form factor of the target nucleus, for which we adopt the dipole nucleon approximation \cite{Perdrisat:2006hj}
\begin{equation}\label{eq:form_factor}
    F_{{\rm SI},{\cal T}}(q)=\left(1+\frac{q^2}{\Lambda^2_{\cal T}}\right)^{-2},
\end{equation}
with  $\Lambda_{\cal T}\simeq 0.843\, \mathrm{GeV}\left(\frac{0.8791\, \rm fm}{R_{\cal T}}\right)$ and $R_{\cal T}$ the charge radius of the particular target nucleus (see e.g. \citep{Angeli:2004kvy} for a comprehensive list of charge radii). Finally, $\sigma^{\rm coh}_{{\rm SI},{\cal T}}$ and $\sigma^{\rm inc}_{{\rm SI},{\cal T}}$ are respectively the coherent and incoherent spin-independent scattering cross sections with the target nucleus ${\cal T}$ at zero momentum transfer, which are related to the DM-proton and DM-neutron cross-section, $\sigma_p$ and $\sigma_n$,  through
\begin{align}
    \sigma^{\rm coh}_{{\rm SI},{\cal T}}&=
   \sigma_p\left(\frac{\mu_{\cal T}}{\mu_p}\right)^2
\left[Z_{\cal T} f^p+(A_{\cal T}-Z_{\cal T})f^n\right]^2, \\
    \sigma^{\rm inc}_{{\rm SI},{\cal T}}&= Z_{\cal T} \sigma_p+(A_{\cal T}-Z_{\cal T})\sigma_n\;,
\end{align}
where $\mu_a=m_a m_\chi/(m_a+m_\chi)$ is the reduced mass of the system of the DM particle and the particle $a$, with $a=p$ for the proton and $a={\cal T}$ for the target nucleus with mass number $A_{\cal T}$ and atomic number $Z_{\cal T}$. The terms $f^{p,n}$ are the contributions of protons and neutrons to the total coupling strength. For an isoscalar interaction ($f^p=f^n=1$, $\sigma_p=\sigma_n$) one obtains
\begin{align}
    \sigma^{\rm coh}_{{\rm SI},{\cal T}}&=\sigma_p\left(\frac{\mu_{\cal T}}{\mu_p}\right)^2A_{\cal T}^2,\\
    \sigma^{\rm inc}_{{\rm SI},{\cal T}}&=\sigma_p A_{\cal T}.
\end{align}

\begin{figure}[tb]
	\centering 
	\includegraphics[width=0.4\textwidth]{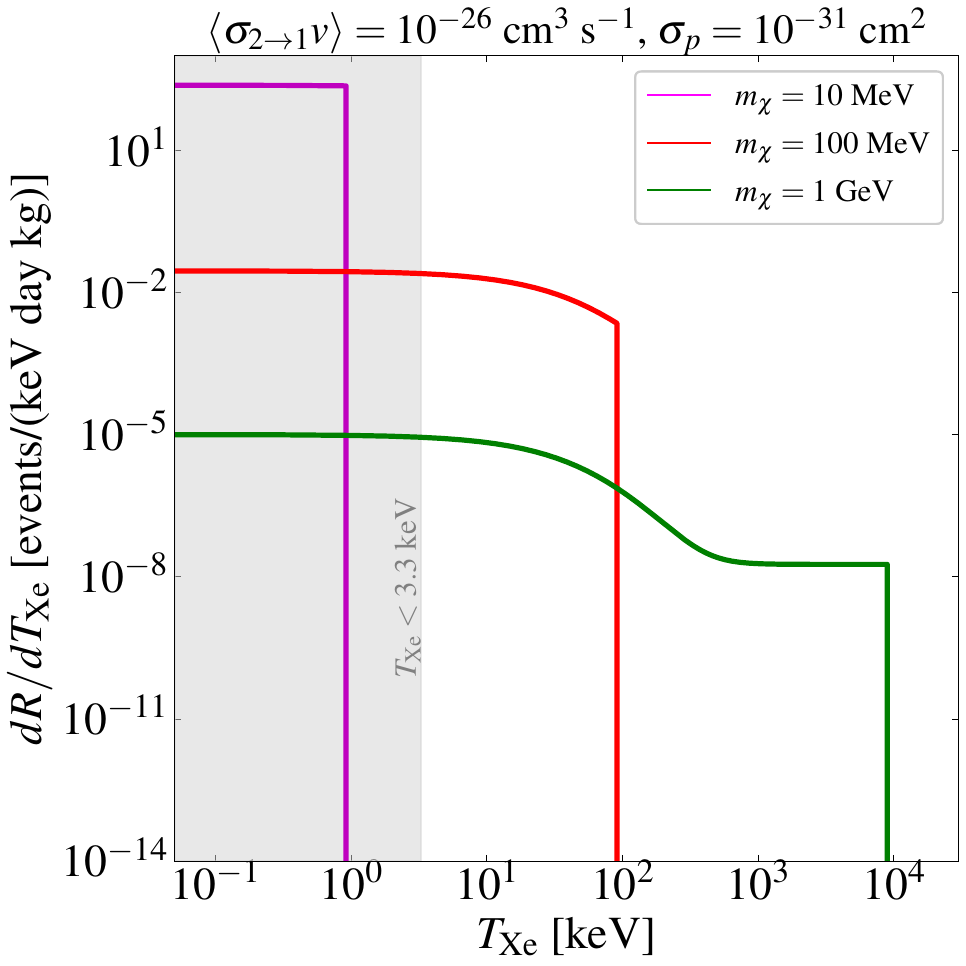}	
	\caption{Recoil spectra for the elastic scattering of DM off $^{131}$Xe nuclei from  semi-annihilation in the Galactic Center with cross-section $\langle \sigma_{2\rightarrow 1} v\rangle=10^{-26}\,{\rm cm}^3/{\rm s}$ and a DM-proton scattering cross-section $\sigma_p=10^{-31}\,\mathrm{cm}^2$.} 
	\label{fig:DifferentialRate}%
\end{figure}

For illustration, we show in Fig.~\ref{fig:DifferentialRate} the recoil spectra of a $^{131}$Xe nucleus after the scattering off of a DM particle with mass $m_\chi=10$ MeV (violet), 100 MeV (red) and 1 GeV (green) assuming an annihilation cross-section for the semi-annihilation process
$\langle\sigma_{2\to1}v\rangle=10^{-26}\,{\rm cm}^3/\rm s$ and a DM-proton scattering cross-section $\sigma_p=10^{-31}~\mathrm{cm}^2$, which is allowed by current direct detection experiments (including the cosmic-ray boosted component~\citep{Bringmann:2018cvk}). The recoil spectra show a characteristic sharp cut-off at $T_{\cal T}^{\rm max}$ given by Eq.~(\ref{eq:T_max})  with $T_\chi=m_\chi/4$, which is qualitatively different to the expected recoil spectra from cosmic-ray boosted DM \citep{Yin:2018yjn,Bringmann:2018cvk,Ema:2018bih}, or blazar-boosted DM \citep{Wang:2021jic}. 
The existence of this cut-off in the recoil spectrum implies a mass threshold for the DM particle, for which the nuclear recoil becomes undetectable. Concretely, this minimum mass is
\begin{equation}\label{eq:min_mass}
    m_\chi^{\rm min}=\frac{(5/4)m_{\cal T}}{\frac{(9/8)m_{\cal T}}{T^{\rm th}}-1}\left[1+\frac{3}{5}\sqrt{1+\frac{2m_{\cal T}}{T^{\rm th}}}\right],
\end{equation}
corresponding to a DM particle with kinetic energy  $T_\chi=m_\chi/4$ producing a nuclear recoil 
with the minimum threshold energy of the experiment $T^{\rm th}$. 
For the case of the XENONnT experiment, which has threshold $T_{\rm XENON}^{\rm th}=3.3$ keV~\citep{XENON:2023cxc}, this minimum mass is 19 MeV. Furthermore, it is noteworthy that the differential rate can be as large as $10^{-2}$ events/(keV day kg) for $m_\chi=100$ MeV at $T_{\rm Xe}\gtrsim 3.3$ keV. This translates into tens of events in a ton-scale xenon detector, showing the high sensitivity of current experiments to semi-annihilation boosted DM from the Galactic Center, which allows to probe new regions of the parameter space of this scenario.

Finally, the total event rate (in units of the number of events$/$(day$\times$kg)) is found by integrating Eq.~\eqref{eq:diff_rate} over the energy window of a given experiment, and over all target nuclei:
\begin{equation}\label{eq:int_scatt_rate}
    R=\sum_{\cal T}\int
    {dT_{\cal T}\,\frac{dR_{\cal T}}{dT_{\cal T}}}.
\end{equation}
The number of expected events, $N_{\rm exp}$, follows from multiplying by the exposure of the experiment. 

\section{Signals at direct detection and neutrino experiments}
\label{sec:direct_detection_constraints}
We now derive limits on the DM-proton scattering cross section $\sigma_p$ from the non-observation of an excess of nuclear recoil events in the DM direct detection experiments XENONnT and CRESST-II, as well as in the neutrino experiment MiniBooNE.~\footnote{ Other possible constraints on boosted dark matter models were listed in \cite{Agashe:2014yua}. These constraints, on the other hand, do not apply to the semi-annihilation scenario under consideration in this work, which contains one neutrino in the final state.} The relevant details about these experiments, as well as our approach to calculate the limits, can be found in Appendix \ref{app:derivation}.
We will focus our analysis on DM masses above 1 MeV since these experiments have no sensitivity to semi-annihilations of lighter DM particles ({\it cf.} Eq.(\ref{eq:min_mass})). Further, we will focus on DM masses below 10 GeV, since this is the region where experiments lose sensitivity to DM particles in the Standard Halo Model; for larger masses, the flux of DM particles at Earth assuming the Standard Halo Model is $\Phi_{\rm SHM}\sim 10^7 (m_\chi/100\,{\rm MeV})^{-1}\,{\rm cm}^{-2}\,{\rm s}^{-1}$, which is many orders of magnitude larger than the one originating from semi-annihilations in the Galactic Center. Thus, above 10 GeV, the sensitivity of experiments does not change significantly by the inclusion of the boosted DM component from semi-annihilations.

\begin{figure}[tb]
	\centering
	\includegraphics[width=0.46\textwidth]{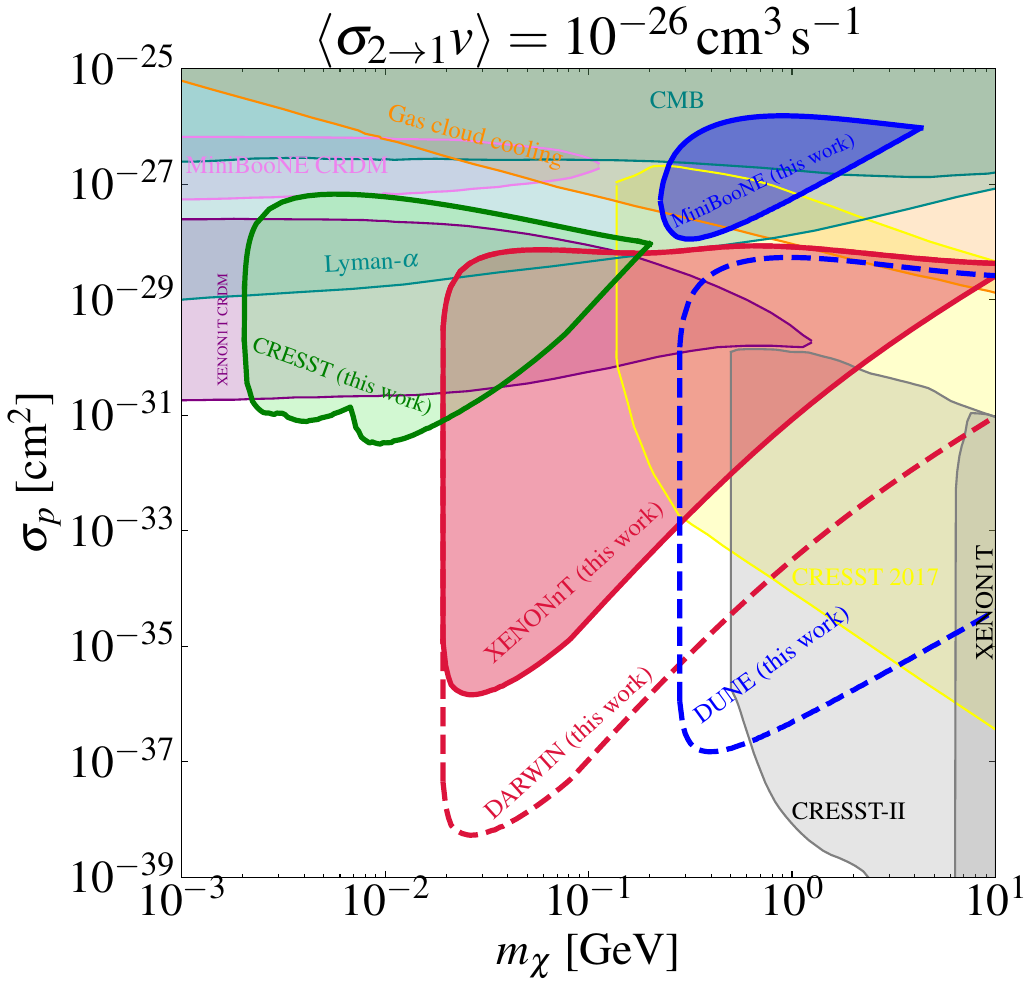}
	\caption{Excluded regions for the spin-independent DM-proton cross section from XENONnT (red region), CRESST (green region), MiniBooNE (blue region) as well as the projected sensitivities for DARWIN (dashed red line)  and DUNE (dashed blue line). We assume a cross-section for the semi-annihilation $\langle \sigma_{2\rightarrow 1}v\rangle=10^{-26}\,{\rm cm}^3\,{\rm s}^{-1}$ and a NFW halo profile.  For comparison, we also include the constraints from XENON1T and CRESST assuming the Standard Halo Model, as well as the constraints from cosmic ray boosted DM (CRDM) \citep{Bringmann:2018cvk}, CMB observations \citep{Xu:2018efh}, Lyman-$\alpha$ \citep{Rogers:2021byl} and gas cloud cooling \citep{Bhoonah:2018wmw}.
 }
	\label{fig:2to1Constraints_sigmav1e-26}
\end{figure}

We show in Fig.~\ref{fig:2to1Constraints_sigmav1e-26} conservative upper limits on the DM-proton scattering cross-section from requiring that the expected number of signal events from semi-annihilations in the Galactic Center is in agreement with the data from CRESST, XENONnT and MiniBooNE, assuming that the cross-section is $\langle \sigma_{2\rightarrow 1} v\rangle=10^{-26}\,{\rm cm}^3\,{\rm s}^{-1}$ (the flux as a function of the mass follows from Eq.~(\ref{eq:Phi_BDM})). We also show in the Figure the constraints from the two CRESST phases~\citep{CRESST:2015txj,CRESST:2017ues} and the XENON1T experiment~\citep{XENON:2017vdw} resulting from assuming that the DM flux at the detector is entirely described by the Standard Halo Model,
from the flux component generated by DM upscattering by cosmic-rays~\citep{Bringmann:2018cvk}, as well as from the potential effect of DM-proton interactions on the cosmic microwave background radiation \citep{Xu:2018efh}, the Lyman-$\alpha$ forest \citep{Rogers:2021byl} and gas cloud cooling \citep{Bhoonah:2018wmw}. We also show our estimated reach in parameter space of the DARWIN and DUNE experiments.

We find that the XENONnT experiment can probe scattering cross-sections as low as $\sigma_p\sim 10^{-36}\,{\rm cm}^2$ for $m_\chi\sim 30$ MeV, the CRESST experiment,  $\sigma_p\sim 10^{-31}\,{\rm cm}^2$ for $m_\chi\sim 3-30$ MeV, and MiniBooNE, $\sigma_p\sim 10^{-28}\,{\rm cm}^2$ for $m_\chi\sim 300$ MeV. The sensitivity of the CRESST experiment is comparable to that of Cosmic Ray boosted DM. In contrast, the sensitivity of the XENONnT experiment is up to five orders of magnitude better. It is also apparent from the Figure the existence for each experiment of a minimum detectable DM mass from the semi-annihilation,  given in Eq.~(\ref{eq:min_mass}), and which is $m_{\chi}^{\rm min}\simeq 20\,\mathrm{MeV}$ for the XENONnT experiment, $m_{\chi}^{\rm min}\simeq 2\,\mathrm{MeV}$ for the CRESST experiment (corresponding to scattering with oxygen), and $m_{\chi}^{\rm min}\simeq 200\,\mathrm{MeV}$ for MiniBooNE. The upcoming DARWIN experiment can reach cross-sections as low as $\sigma_p\sim 10^{-38}\,{\rm cm}^2$ for $m_\chi\sim 30$ MeV and DUNE as low as $\sigma_p\sim 10^{-37}\,{\rm cm}^2$ for $m_\chi\sim 300$ MeV.

The plot also reflects the loss of sensitivity of these experiments for a very large cross-section, due to the attenuation of the DM flux by interactions with the rock when DM particles traverse the Earth's crust. The rate of energy loss of a DM particle as it traverses a distance $z$ in a medium filled with targets ${\cal T}$ with constant number density $n_{\cal T}$ reads:
\begin{align}
\label{eq:energy_loss_full}
    \frac{dT_\chi }{dz}=-\sum\limits_{\cal T}{n_{\cal T}}
    \int_0^{T_{\cal T}^{\rm max}}dT_{\cal T}\,T_{\cal T}\,
     \frac{d\sigma_{\chi{\cal T}}}{dT_{\cal T}}(T_\chi,T_{\cal T}),
\end{align}
where $T^{\rm max}_{\cal T}$ and $d\sigma_{\chi{\cal T}}/dT_{\cal T}$ are given in Eqs.(\ref{eq:T_max}) and (\ref{eq:diff_cross_section}) respectively. For $m_\chi\ll m_{\cal T}$, the rate of energy loss can be conservatively bounded by~\citep{Bringmann:2018cvk}
\begin{align}
\label{eq:energy_loss_full_2}
    \frac{dT_\chi }{dz}\gtrsim-\frac{T_\chi^2+2m_\chi T_\chi}{2m_\chi \ell},
\end{align}
which corresponds to the approximation that all targets are point-like, and where $\ell$ is a characteristic length given by
\begin{align}
    \ell^{-1}=\sigma_p\sum\limits_{\cal T}{2n_{\cal T}} A_{\cal T}^2\left(\frac{m_\chi}{m_{\cal T}}\right)\left(1+\frac{m_\chi}{m_p}\right)^2\left(1+\frac{m_\chi}{m_{\cal T}}\right)^{-4}\equiv \sigma_p {\cal V}^{-1},
\end{align}
with ${\cal V}$ a calculable constant with units of volume.
The solution to Eq.~(\ref{eq:energy_loss_full_2}) reads:
\begin{align}
T_\chi(z)&=\frac{2m_\chi T_\chi(0)\mathrm{e}^{-z/\ell}}{2m_\chi+T_\chi(0)-T_\chi(0)\mathrm{e}^{-z/\ell}}\cr
&\hspace{-0.1cm}\gtrsim T_\chi(0) e^{-z/\ell}\left[1-\frac{T_\chi(0)}{2m_\chi}\left(1-e^{-z/\ell}\right)
\right]\approx T_{\chi}(0)e^{-z/\ell},
\end{align}
where in the approximation in the last step we have used that $T_\chi(0)=m_\chi/4$. The DM particles will be unobservable at the detector if the kinetic energy of the recoiling nucleus $T_{\cal T}$ produced by a DM particle after traversing a depth $z_{\rm exp}$ (corresponding to the location of the experiment) is smaller than the threshold of the experiment $T_{\rm th}$, for all targets in the detector. This translates into the condition for the DM kinetic energy
$T_\chi(z_{\rm exp})<{\rm min}_{\cal T}\{ T_{\chi,{\cal T}}^{\rm min}(T^{\rm th})\}\equiv T_\chi^{\rm min}$ where $T_{\chi,{\cal T}}^{\rm min}(T_{\cal T})$ is given in Eq.~(\ref{eq:T_min}). In turn, this condition implies an upper limit on the DM-proton cross-section
\begin{align}\label{eq:sigma_atten}
\sigma_p \gtrsim \frac{\cal V}{z_{\rm exp}}\ln\left(\frac{m_\chi}{4T_\chi^{\rm min}}\right),
\end{align}
above which DM particles become undetectable in the corresponding underground experiment. One should note that the column density traversed by DM particles is time-dependent, as the galactic center moves in the sky as viewed at the experimental location. For instance, at the Gran Sasso National Laboratory, where the XENON and CRESST experiments are located, the altitude of the Galactic Center varies between $-76.5^\circ$ and $18.5^\circ$, which amounts to a path of 12400 km and 1.4 km through Earth. Since we are interested in deriving conservative lower limits on the cross-section we will adopt the shortest distance, $z_{\rm exp}=1.4$ km, and a constant density  $\rho=2.71\,{\rm g}/{\rm cm}^3$ of rock composed of 48\% of oxygen, 30\% of calcium, 12\% of carbon and 5.6\% of magnesium (as well as traces of other elements) \cite{Wulandari:2003cr}; for MiniBooNE, we adopt $z_{\rm exp}=6$ m, and for DUNE, $z_{\rm exp}=1.5$ km.

The lower limits on the cross-section from the XENONnT, CRESST and MiniBooNE are shown in Fig.~\ref{fig:2to1Constraints_sigmav1e-26}, and were obtained by solving numerically Eq.~(\ref{eq:energy_loss_full}), without further approximations, and imposing $T_\chi(z_{\rm exp})< T_\chi^{\rm min}$. In our analysis, we have found that the inclusion of the incoherent effects in the scattering modifies significantly the lower limits on the cross-section for GeV mass DM. Our limits could be refined taking into account the motion of the Galactic Center in the sky relative to the location of the experiment, as well as the non-constant density of the Earth, which would translate into a larger excluded region. These enlarged excluded regions are nevertheless excluded already by other experiments, and therefore we will not pursue here this improved approach. 

\section{Conclusions}
\label{sec:conclusions}

We have considered the scenario where DM particles semi-annihilate, producing one DM particle and one Standard Model particle in the final state. In this scenario, the DM particle produced in the semi-annihilation necessarily has a larger momentum than the incoming particles. It could therefore produce a component in the DM flux at Earth with speeds larger than the escape velocity of the Milky Way, opening the possibility of detecting in direct detection experiments DM particles with mass in the sub-GeV range. 

We have derived constraints on the DM-proton cross-section using data from the experiments XENONnT, CRESST-II and MiniBooNE. For a typical semi-annihilation cross-section of $\langle\sigma_{2\rightarrow 1} v\rangle =10^{-26}\,{\rm cm}^3\,{\rm s}^{-1}$, which leads via thermal freeze-out to a DM abundance in the ballpark of the observed value, we find that the limits on the DM-proton cross-section from the XENONnT experiment are about five orders of magnitude better than existing limits in the mass range $m_{\chi}\sim 20-300$ MeV, while the limits from the CRESST experiment are slightly better in mass range $2-30\,{\rm MeV}$. The limits from MiniBooNE do not rule out uncharted parameter space, although are complementary to other existing limits. The upcoming DARWIN and DUNE experiments will close in on the parameter space of DM models, reaching respectively a cross-section $\sigma_p\sim 10^{-38}\,{\rm cm}^2$ for $m_\chi\sim 30$ MeV and $\sigma_p\sim 10^{-37}\,{\rm cm}^2$ for $m_\chi\sim 300$ MeV.

\section*{Acknowledgements}
We thank Torsten Bringmann for instructing how to use DarkSUSY for monochromatic boosted DM fluxes and for useful comments on the results. MF acknowledges the Mainz Institute for Theoretical Physics (MITP) of the Cluster of Excellence PRISMA+ (Project ID 390831469) for receiving feedback at the last stage of this collaboration during ``The Dark Matter Landscape: From Feeble to Strong Interactions''. The work of BBK, MF, and AI is supported by the Deutsche Forschungsgemeinschaft (DFG, German Research Foundation) under Germany’s Excellence Strategy - EXC-2094 - 390783311 and under the Collaborative Research Center SFB1258 grant - SFB-1258 - 283604770. The work of TT is supported by JSPS KAKENHI Grant Number 23H04004.

\appendix

\section{Derivation of limits}
\label{app:derivation}
In this Appendix, we briefly describe our methodology to calculate conservative limits on the cross-section for each experiment.
\subsection*{XENONnT}

We impose that the event rate induced by semi-annihilations does not exceed the experimental upper limit, $R<R_{\rm exp}$, with $R$ 
given in Eq.~(\ref{eq:int_scatt_rate}) with an energy window of interest $[3.3,60.5]$ keV~\citep{XENON:2023cxc}.
We determined the experimental upper limit on the rate of nuclear recoils following \cite{Bringmann:2018cvk}, translating the upper limit on the spin-independent cross-section reported by an experimental collaboration into an upper limit on the event rate, utilizing the fact that the experimental collaborations assume the Standard Halo Model in their analysis. One obtains
\begin{equation}
    R_{\rm exp}=\frac{\kappa\, v_c\, \rho_\chi}{m_{\cal T}}\,A^2\left(\frac{m_{\cal T}}{m_p}\right)^2\left(\frac{\sigma_{p}^{\rm lim}}{m_\chi}\right)_{m_\chi\gg m_{\cal T}},
\end{equation}
where $\kappa\approx 0.37$ and $\sigma_{p}^{\rm lim}/m_\chi\simeq 6.1\times10^{-49}\,\mathrm{cm}^2/\mathrm{GeV}$ ({\it cf.} Fig.~4 of \citep{XENON:2023cxc}).

\subsection*{CRESST-II}

We require that the number of expected events is smaller than the number of observed ones, $N_{\rm exp}<N_{\rm obs}$, with $N_{\rm exp}$ given under Eq.~(\ref{eq:int_scatt_rate}), over the energy range $[0.307,40]\,\rm keV$~\citep{CRESST:2015txj}. The collaboration observed $N_{\rm obs}=1949$ events within the acceptance region for exposure of $52.15\, \mathrm{kg}\,\mathrm{days}$ \citep{CRESST:2017fmc}, from where we derive a conservative upper limit of the event rate from DM scatterings, mirroring the procedure of the CRESST collaboration \citep{Angloher:2016jsl}. We assume scattering of DM with $\mathrm{CaWO}_4$ crystals.

\subsection*{MiniBooNE}

We impose that the DM induced scattering rate per proton, calculated from  Eq.~(\ref{eq:int_scatt_rate}) with protons as the target, satisfies $\Gamma_p<1.5\times10^{-32}\,\mathrm{s}^{-1}$ \citep{Bringmann:2018cvk} in the energy window $[35\,\mathrm{MeV},1.4\,\mathrm{GeV}]$ \citep{MiniBooNE:2020pnu, Bringmann:2018lay}.

\subsection*{DARWIN}

The planned DARWIN experiment will also use xenon as the target and will have the same location as the XENON experiment. To calculate the sensitivity in the cross-section we will simply scale the current limits from the XENONnT experiment by a factor 270 \citep{DARWIN:2016hyl}.

\subsection*{DUNE}
To estimate the DUNE sensitivity, we impose the number of annihilation-boosted signals to be smaller than the number of signals $N_{\rm sig}$ at a significance of $S=2$ with $S=N_{\rm sig}/\sqrt{N_{\rm sig}+N_{\rm bkg}}$ for a 40 kton detector and 10 years of exposure with an energy range of $[50\,\mathrm{MeV},10\,\mathrm{GeV}]$ \citep{DUNE:2015lol}. We calculate the number of neutrino background events as \citep{Aoki:2023tlb}
\begin{equation}
    N_{\rm bkg}=N_N T\sum_\alpha{\int{\frac{\sigma_{\nu_\alpha N}}{A_{\rm Ar}} \frac{d^2\Phi_{\nu_\alpha}}{d E_{\nu_\alpha}d\Omega}dE_{\nu_\alpha}d\Omega}},
\end{equation}
where $\alpha$ runs over electron and muon neutrinos (and antineutrinos), $A_{\rm Ar}=39.9$ is the argon mass number, $N_N=2.41\times10^{34}$ is the number of nucleons in a 40 kton volume of liquid argon and $T=10$ years.
The DUNE detector will be located at the Sanford Underground Research Laboratory in Lead, South Dakota, therefore we use the differential atmospheric neutrino fluxes $d^2\Phi_{\nu_\alpha}/d E_{\nu_\alpha}d\Omega$ at the Homestake experiment based on the HAKKM2014 model\footnote{Tables are found in http://www-rccn.icrr.u-tokyo.ac.jp/mhonda/public/nflx2014/index.html.} \citep{Honda:2015fha} and average the fluxes with the minimum and maximum solar modulation effect. We have obtained the relevant neutral current neutrino-argon cross sections from the neutrino event generator GENIE \citep{Andreopoulos:2009rq} and we use the solid angle $\Delta
\Omega = 0.668$ to get a conservative DUNE sensitivity \citep{Aoki:2023tlb} to obtain $N_{\rm bkg}=1952$ background events.

\bibliographystyle{elsarticle-num} 
\bibliography{bibliography}






\end{document}